\documentclass[sigconf,natbib=true]{acmart}
\usepackage{multirow}
\usepackage{graphics}
\usepackage{arydshln}

\usepackage{colortbl}
\usepackage{graphicx}
\usepackage[table]{xcolor}
\usepackage{diagbox}
\usepackage{tabularx}              % 자동 열 너비 조절을 위해
\usepackage{booktabs}              % 깔끔한 테이블 선을 위해
\usepackage{array}                 % 열 정렬을 위해
\usepackage{kotex}[cjk]
\usepackage{subcaption}
\usepackage{geometry}
\usepackage{tabularx} % 폭 자동 조절을 위해 필요

\AtBeginDocument{%
  }
\copyrightyear{2026}
\acmYear{2026}
\setcopyright{cc}
\setcctype{by}
\acmConference[SIGIR '26]{Proceedings of the 49th International ACM SIGIR Conference on Research and Development in Information Retrieval}{July 20--24, 2026}{Melbourne, VIC, Australia}
\acmBooktitle{Proceedings of the 49th International ACM SIGIR Conference on Research and Development in Information Retrieval (SIGIR '26), July 20--24, 2026, Melbourne, VIC, Australia}
\acmDOI{10.1145/3805712.3809843}
\acmISBN{979-8-4007-2599-9/2026/07}

\begin{document}

\title{Beyond Hard Negatives: The Importance of Score Distribution in Knowledge Distillation}

\author{Youngjoon Jang}
\orcid{0009-0005-3591-638X}
\affiliation{%
  \institution{Korea University}
  \city{Seoul}
  \country{South Korea}
}
\email{dew1701@korea.ac.kr}

\author{Seongtae Hong}
\orcid{0009-0002-2073-7731}
\affiliation{%
  \institution{Korea University}
  \city{Seoul}
  \country{South Korea}
}
\email{ghdchlwls123@korea.ac.kr}

\author{Hyeonseok Moon}
\orcid{0000-0002-0841-4262}
\authornote{Corresponding author}
\affiliation{%
  \institution{Korea University}
  \city{Seoul}
  \country{South Korea}}
\email{glee889@korea.ac.kr}

\author{Heuiseok Lim}
\orcid{0000-0002-9269-1157}
\authornotemark[1]
\affiliation{%
  \institution{Korea University}
  % \institution{Human-inspired AI Research}
  \city{Seoul}
  \country{South Korea}}
\email{limhseok@korea.ac.kr}

\renewcommand{\shortauthors}{Jang et al.}

% \begin{abstract}
% Transferring knowledge from a cross-encoder teacher via Knowledge Distillation (KD) has become a standard paradigm for training dense retrieval models. However, existing studies have overlooked the systematic composition of training data, leaving the impact of the teacher’s score distribution underexplored. In this work, we demonstrate that such oversight prevents the student from learning the full score distribution of the teacher, consequently hampering generalization performance. To effectively emulate the teacher's distribution, we propose a stratified sampling strategy that uniformly covers the entire score spectrum. Experiments on in-domain and out-of-domain benchmarks confirm that stratified sampling, which preserves the variance and entropy of teacher scores, significantly outperforms competing methods including top-K and random sampling. These findings suggest that the essence of distillation lies not merely in distinguishing hard negatives, but in preserving the global ranking geometry by learning the diverse semantic distances between queries and documents as perceived by the teacher.
% \end{abstract}
\begin{abstract}
Transferring knowledge from a cross-encoder teacher via Knowledge Distillation (KD) has become a standard paradigm for training retrieval models. While existing studies have largely focused on mining hard negatives to improve discrimination, the systematic composition of training data and the resulting teacher score distribution have received relatively less attention. In this work, we highlight that focusing solely on hard negatives prevents the student from learning the comprehensive preference structure of the teacher, potentially hampering generalization. To effectively emulate the teacher score distribution, we propose a Stratified Sampling strategy that uniformly covers the entire score spectrum. Experiments on in-domain and out-of-domain benchmarks confirm that Stratified Sampling, which preserves the variance and entropy of teacher scores, serves as a robust baseline, significantly outperforming top-K and random sampling in diverse settings. These findings suggest that the essence of distillation lies in preserving the diverse range of relative scores perceived by the teacher.
\end{abstract}

%    <concept>
%        <concept_id>10002951.10003317.10003338</concept_id>
%        <concept_desc>Information systems~Retrieval models and ranking</concept_desc>
%        <concept_significance>500</concept_significance>
%        </concept>
%    <concept>
%        <concept_id>10002951.10003317.10003338.10003341</concept_id>
%        <concept_desc>Information systems~Language models</concept_desc>
%        <concept_significance>500</concept_significance>
%        </concept>
%    <concept>
%        <concept_id>10002951.10003317.10003371.10003381.10003385</concept_id>
%        <concept_desc>Information systems~Multilingual and cross-lingual retrieval</concept_desc>
%        <concept_significance>500</concept_significance>
%        </concept>
%  </ccs2012>

% \ccsdesc[500]{Information systems~Language models}
% \ccsdesc[500]{Information systems~Retrieval models and ranking}
% \ccsdesc[500]{Information systems~Multilingual and cross-lingual retrieval}

\begin{CCSXML}
<ccs2012>
   <concept>
       <concept_id>10002951.10003317.10003338</concept_id>
       <concept_desc>Information systems~Retrieval models and ranking</concept_desc>
       <concept_significance>500</concept_significance>
       </concept>
 </ccs2012>
\end{CCSXML}

\ccsdesc[500]{Information systems~Retrieval models and ranking}

\keywords{Information Retrieval, Dense Retrieval, Data Diversity}

\maketitle

\section{INTRODUCTION}

Dense retrieval models have established themselves as key components in large-scale Information Retrieval (IR) systems due to their efficiency and scalability~\cite{ir1,ir2,ir3,ir4}. However, due to the structural constraint of compressing document semantics into a single vector, they face limitations in ranking quality compared to models that utilize richer contextual interactions~\cite{reranker1,reranker4,colbertv1,colbertv2,jinacolbert,clavie2025jacolbertv2}. To bridge the gap between efficiency and performance, Knowledge Distillation~(KD), which trains a dense retriever as a student using a powerful yet computationally expensive model as a teacher, is widely adopted~\cite{distil1,topk2,distil4}. In this process, the signal provided by the teacher is not binary relevance but rather numerical values reflecting relative preference over a set of candidate documents, and the student is trained to approximate this using loss functions such as KL divergence or MarginMSE.

Nevertheless, the composition of training samples for distillation has been explored primarily through the lens of mining difficult examples. Common practices often rely on heuristic selections, such as fixing the top-K hard negatives mined by a first-stage retriever or simply utilizing random samples~\cite{distil5,topk1,topk2,topk3,topk4}. However, these heuristic compositions may result in observing only a limited or biased segment of the teacher's score distribution. Consequently, the model may struggle to learn decision boundaries of varying difficulty levels. In other words, the critical question regarding distillation data composition needs to be redefined: rather than relying on heuristic candidate mining, it should focus on whether diverse distribution of the teacher's score is sufficiently sampled.

This study systematically investigates this issue from the perspective of data composition and proposes Stratified Sampling to preserve the diversity of the teacher score distribution. The proposed method uniformly places quantile anchors across the score distribution of candidates and selects the candidates closest to each anchor score, ensuring score coverage and distributional representativeness. To isolate the effect of score distribution, we construct a controlled candidate pool combining top-retrieved and random documents from MSMARCO~\cite{msmarco}, and compare $k$ candidates selected via different sampling strategies within an identical training pipeline. We intentionally use this fixed pool to strictly isolate the impact of score distribution from the confounding variables of complex dynamic miners.

We design the experiments with a two-stage training process to decouple the effects of the model and the objective function. First, pretrained encoder models are adapted via Contrastive Learning~(CL). Subsequently, triplets sampled based on reranker scores are trained using KL-Divergence~\cite{kldiv} and MarginMSE~\cite{marginmse}. Experimental results demonstrate that Stratified Sampling consistently achieves robust performance across all base models and objective functions. Notably, we observe that Stratified Sampling remains robust even under varying numbers of candidates.

This paper empirically demonstrates that simple Stratified Sampling can simultaneously improve both in-domain and out-of-domain performance without complex curriculum scheduling. This presents a practical, robust alternative to heuristic data composition methods and offers a standard criterion for future distillation data design.

\section{RELATED WORKS}
While bi-encoder based dense retrieval is widely employed for efficient first-stage retrieval, a performance gap remains compared to cross-encoder rerankers~\cite{reranker2,reranker3,reranker_bert,reranker_cedr,reranker_t5}. To bridge this gap, Knowledge Distillation (KD) has been actively researched. Representative approaches involve mimicking the continuous scores using distribution matching based on KL divergence or regression objectives such as MarginMSE~\cite{distil5,marginmse}. 

Meanwhile, negative selection is a critical factor determining performance. Various candidate generation techniques have been proposed, including hard negative mining~\cite{topk1,topk2,huang2024pairdistill} and denoising strategies~\cite{rocketqav1,distil5}. In the context of distillation, strategies constructing training samples based on simple heuristics (in-batch, BM25, random) are widely adopted~\cite{distil4,simlm,coder}. 

% --- [수정된 문단 시작] ---
Recently, advanced strategies have been proposed to improve generalization, ranging from geometric constraints~\cite{kim2023embeddistill} and adaptive dark example selection~\cite{tao2024adam} to curriculum learning~\cite{prod, zeng2022curriculum}. While effective, these methods often require complex scheduling, auxiliary losses, or dynamic sampling pipelines. In contrast, our work focuses on the static composition of the training data itself. We propose Stratified Sampling as a simpler, parameter-free alternative that statistically ensures the representativeness of the teacher's score distribution without the need for dynamic adjustments.

\section{EXPERIMENTAL SETUP}
In this study, we conduct experiments using various combinations of student backbones (bert-base-uncased\footnote{\url{https://huggingface.co/google-bert/bert-base-uncased}}, distilbert-base-uncased\footnote{\url{https://huggingface.co/distilbert/distilbert-base-uncased}}, co-condenser-marco\footnote{\url{https://huggingface.co/Luyu/co-condenser-marco}}) and distillation objective functions (KLDiv, MarginMSE). However, the primary variable for comparison is the data composition, specifically the document candidate sampling strategy (retriever-top, reranker-top, mid, low, random, stratified).
The full information of the training and evaluation details are presented in Table~\ref{tab:setup}.

\begin{table}[h]
\centering
\small
\caption{The overview of experimental setup.}
\label{tab:setup}
\renewcommand{\arraystretch}{1.04} % 줄바꿈이 들어가면 행간이 좁아 보이므로 1.2 정도로 살짝 늘리는 것을 추천합니다.

\begin{tabular}{l|l}
% \noalign{\hrule height 1.2pt} 
\toprule

% 섹션 1
\multicolumn{2}{l}{\textbf{Modeling and Training Objectives:}} \\
Models & bert, distilbert, co-condenser \\
Training Objectives & \begin{tabular}[c]{@{}l@{}}1. CL \\ 2. KD (KL-divergence, MarginMSE)\end{tabular} \\
\midrule

% 섹션 2
\multicolumn{2}{l}{\textbf{Training and Evaluation Datasets:}} \\
Training Data & MS MARCO (8.8M docs, 532K queries) \\
1st stage retriever & Qwen3-Embedding-8B \\
2nd stage reranker & Qwen3-Reranker-8B \\
Sampling Strategy & \begin{tabular}[c]{@{}l@{}} retriever-top, reranker-top, mid, \\ low, random, stratified \end{tabular} \\
In-domain Evaluation & MS MARCO-Dev, TREC DL 19 \\
Out-of-domain Evaluation & BEIR (13) \\
\midrule

% 섹션 3
\multicolumn{2}{l}{\textbf{Training Hyperparameters:}} \\
Batch Size & 1024 (CL), 16 (KD) \\ 
Max Length & 512 \\
Pooling Mode & mean \\
Epoch & 1 (CL), 1 (KD) \\
LR & 2e-5 \\
\midrule

% \noalign{\hrule height 1.2pt}
\end{tabular}
\end{table}

\subsection{Data Construction and Sampling}
\paragraph{Candidate Mining}
We utilize MS MARCO-Train for training data. To strictly isolate the impact of score distribution from the retrieval capability of a specific first-stage model, we design a controlled candidate pool. Specifically, for each query, we use Qwen3-Embedding-8B\footnote{\url{https://huggingface.co/Qwen/Qwen3-Embedding-8B}} to mine the top 100 documents. We construct a fixed pool of 200 negatives per query by taking the top-100 documents retrieved by Qwen3-Embedding-8B and sampling an additional 100 documents uniformly at random from the corpus excluding those top-100. This enlarges the difficulty range of negatives and provides a controlled testbed for comparing sampling strategies.
We then run Qwen3-Reranker-8B on the 201 documents (including positive) to obtain teacher scores.

\begin{figure}[t]
    \centering
    \includegraphics[width=\columnwidth]{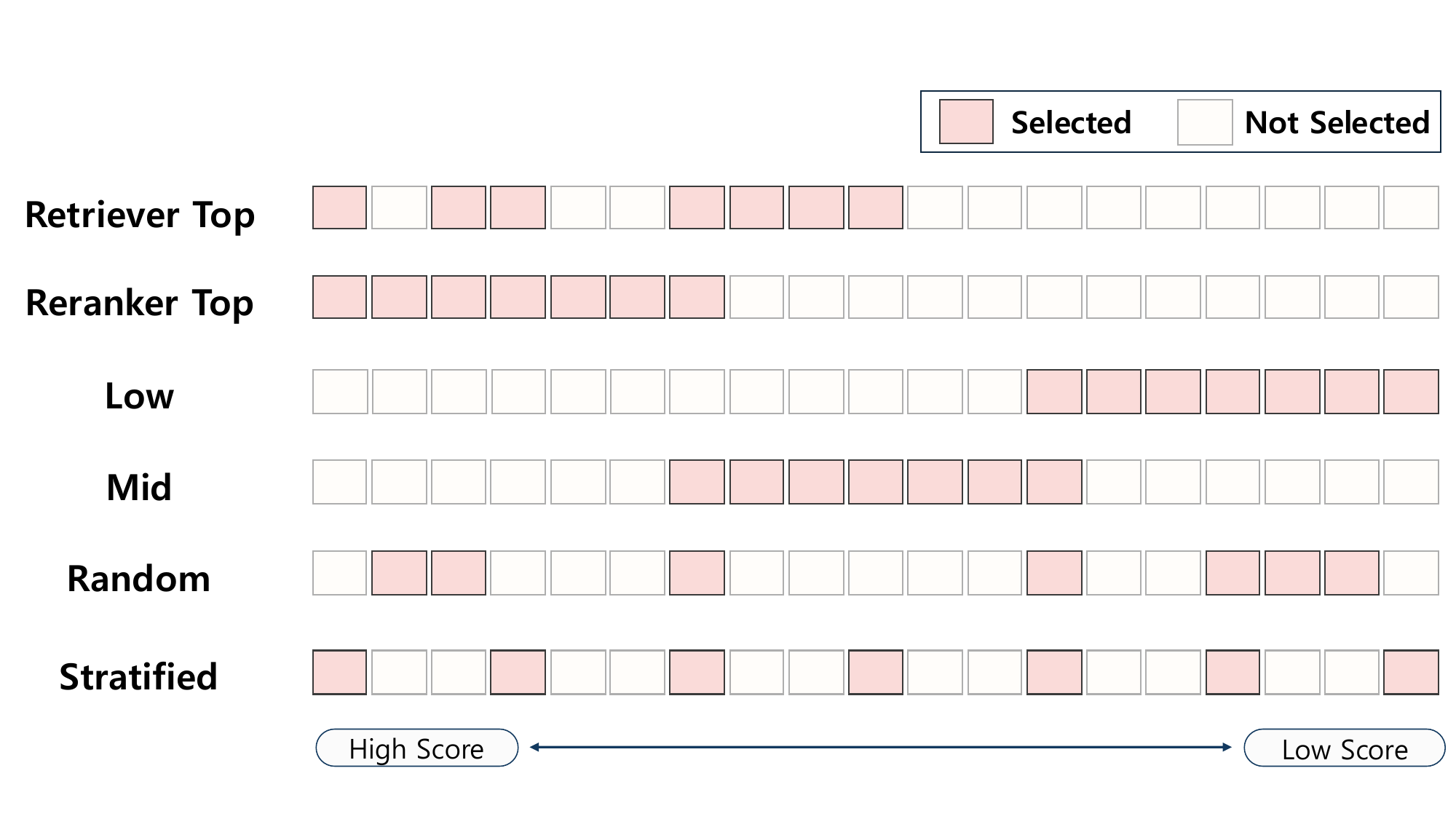}
    \caption{Illustration of candidate selection across different sampling strategies based on the teacher's score distribution.}
    \label{fig:sampling_strategies}
\end{figure}

\paragraph{Sampling}
From the 200 negative candidate documents, we finally select $K$ documents (default $K{=}8$) to construct training samples in the form of $\langle q, d^{+}, \{d^{-}\}_{1:K}\rangle$.
For each query, we apply query-level min–max normalization to the teacher scores over the entire 201-document set (including the positive), yielding $\tilde{s}_i^{(t)}\in[0,1]$. We then perform sampling decisions only over the 200 negatives using these normalized scores $\tilde{s}_i^{(t)}$.
We compare the following six sampling strategies, as illustrated in Figure~\ref{fig:sampling_strategies}:
\begin{description}
  \setlength{\itemsep}{0pt}
  \setlength{\parskip}{0pt}
  \item[\textit{retriever-top}] Top $K$ based on the retriever's mining order.
  \item[\textit{reranker-top}] Top $K$ based on normalized teacher scores.
  \item[\textit{low}] Bottom $K$ based on normalized teacher scores.
  \item[\textit{mid}] $K$ documents around the median of the teacher scores.
  \item[\textit{random}] Random selection of $K$ documents.
  \item[\textit{stratified}] $K$ documents based on quantiles of the teacher scores.
\end{description}

\begin{table*}[t!]
\centering
\caption{Retrieval performance of different sampling strategies across student backbones and distillation objectives.}
\label{tab:main}
\renewcommand{\arraystretch}{1.0} 

% 구조: Model(1) + Sampling(1) | KLDiv(4) | MarginMSE(4)
\begin{tabular}{c | l | c c c c | c c c c}
\toprule
\multirow{3}{*}{\textbf{Model}} & \multicolumn{1}{c|}{\multirow{3}{*}{\textbf{\shortstack{Sampling \\ Strategy}}}} & \multicolumn{4}{c|}{\textbf{KL-divergence}} & \multicolumn{4}{c}{\textbf{MarginMSE}} \\
\cmidrule{3-10}
 & & \multicolumn{2}{c}{\textbf{MSMARCO Dev}} & \textbf{TREC-19} & \textbf{BEIR-13} & \multicolumn{2}{c}{\textbf{MSMARCO Dev}} & \textbf{TREC-19} & \textbf{BEIR-13} \\
 & & \small mrr@10 & \small R@1k & \small nDCG@10 & \small nDCG@10 & \small mrr@10 & \small R@1k & \small nDCG@10 & \small nDCG@10 \\
\midrule

% ==================== BERT-BASE ====================
\multirow{6}{*}{\shortstack{bert-base}} 
 & retriever-top & .166 & .783 & .381 & .217 & .011 & .095 & .045 & .087 \\
 & low & .149 & .722 & .283 & .208 & .192 & .818 & .390 & .264 \\
 & mid & .213 & .764 & .412 & .215 & .236 & .894 & .452 & .314 \\
 & reranker-top & .142 & .689 & .310 & .183 & .053 & .411 & .113 & .115 \\
 & random & .243 & .904 & .458 & .289 & .246 & .913 & .443 & .308 \\
 & stratified & \textbf{.266} & \textbf{.921} & \textbf{.482} & \textbf{.314} & \textbf{.262} & \textbf{.924} & \textbf{.468} & \textbf{.318} \\
\midrule

% ==================== DISTILBERT-BASE ====================
\multirow{6}{*}{\shortstack{distilbert-base}} 
 & retriever-top & .182 & .785 & .417 & .231 & .082 & .548 & .231 & .134 \\
 & low & .150 & .729 & .346 & .195 & .196 & .817 & .438 & .252 \\
 & mid & .150 & .609 & .297 & .201 & .237 & .900 & .470 & \textbf{.338} \\
 & reranker-top & .172 & .684 & .391 & .220 & .069 & .539 & .140 & .129 \\
 & random & .236 & .902 & .479 & .295 & .248 & .916 & .502 & .324 \\
 & stratified & \textbf{.255} & \textbf{.918} & \textbf{.511} & \textbf{.313} & \textbf{.256} & \textbf{.918} & \textbf{.519} & .324 \\
\midrule

% ==================== COCONDENSER ====================
\multirow{6}{*}{\shortstack{co-condenser}} 
 & retriever-top & .276 & .910 & .546 & .319 & .150 & .765 & .324 & .170 \\
 & low & .178 & .788 & .401 & .236 & .251 & .893 & .484 & .329 \\
 & mid & .290 & .925 & .580 & .312 & .282 & .949 & .563 & \textbf{.376} \\
 & reranker-top & .255 & .849 & .518 & .334 & .006 & .014 & .043 & .123 \\
 & random & .295 & .951 & .566 & .326 & .287 & .953 & .564 & .349 \\
 & stratified & \textbf{.313} & \textbf{.959} & \textbf{.587} & \textbf{.365} & \textbf{.307} & \textbf{.961} & \textbf{.580} & \textbf{.376} \\
\bottomrule

\end{tabular}
\end{table*}

\paragraph{Stratified Sampling Details}
To strictly preserve the distributional shape of the teacher's preferences, we implement a deterministic quantile-based Stratified Sampling strategy. Specifically, for a desired number of negatives $K$, we first compute $K$ quantile anchors $\tau_j$ corresponding to evenly spaced cumulative probabilities $p_j = \frac{j-1}{K-1}$ (for $j=1 \dots K$) derived from the min-max normalized teacher scores of the candidate pool. For each anchor $\tau_j$, we iteratively select the distinct candidate document $d^-$ that minimizes the absolute difference $|\tilde{s}_{d^-} - \tau_j|$, thereby ensuring that the training samples structurally mimic the skeleton of the teacher's score distribution without the variance introduced by random sampling.

\subsection{Training}
We propose a two-stage training protocol to conduct independent experiments for distillation.

\paragraph{Contrastive Learning}
In the first stage, we adapt the MLM-based backbone models on MS MARCO-Train using contrastive learning with only in-batch negatives (without negatives). Specifically, for a query $q$, a positive document $d^{+}$, and in-batch negatives $\{d^{-}\}$, the InfoNCE objective function is defined as follows:
\begin{equation*}
\mathcal{L}_{\mathrm{InfoNCE}}= -\log \frac{\exp\left(\mathrm{sim}(\mathbf{h}_q,\mathbf{h}_{d^{+}})/\tau\right)}{\exp\left(\mathrm{sim}(\mathbf{h}_q,\mathbf{h}_{d^{+}})/\tau\right)+\sum_{d^{-}}\exp\left(\mathrm{sim}(\mathbf{h}_q,\mathbf{h}_{d^{-}})/\tau\right)}
\end{equation*}
Here, $\mathbf{h}_q$ and $\mathbf{h}_d$ denote the query and document embeddings, respectively, $\mathrm{sim}(\cdot,\cdot)$ is the similarity function (cosine similarity), and $\tau$ represents the temperature.

\paragraph{Knowledge Distillation}
In the second stage, we perform distillation using the reranker teacher scores.
For each query, given one positive and a set of $K$ candidates $\mathcal{C}=\{d_i\}_{i=1}^{K+1}$, we denote the student score as $s(q,d_i)$ and the teacher score as $t(q,d_i)$.
Consistent with our sampling strategy, we utilize the query-level min-max normalized teacher scores $t(q,d_i) \in [0,1]$ for all distillation objectives to ensure numerical stability and scale consistency.

The KL divergence objective function for listwise distribution matching is defined as:
\begin{equation*}
\begin{aligned}
\mathcal{L}_{\mathrm{KL}} &= \sum_{d_i \in \mathcal{C}} p_i^{t} \log \frac{p_i^{t}}{p_i^{s}},
% \text{where} 
\;
\quad p_i^{\phi} &= \frac{\exp(\phi(q,d_i)/\tau)}{\sum_{d_j \in \mathcal{C}}\exp(\phi(q,d_j)/\tau)} 
% \; \phi \in \{s, t\}
\end{aligned}
\end{equation*}
where we set the temperature $\tau=1.0$ in our experiments.

Additionally, the MarginMSE objective function, which regresses the relative margin derived by the teacher, is defined as follows.
For a positive $d^{+}$ of query $q$ and $K$ candidate documents $\{d_k^{-}\}_{k=1}^{K}$,
\begin{equation*}
\begin{aligned}
\mathcal{L}_{\mathrm{MSE}}
&=\frac{1}{K}\sum_{k=1}^{K}\Bigl(
\bigl(s(q,d^{+})-s(q,d_k^{-})\bigr)
-\bigl(t(q,d^{+})-t(q,d_k^{-})\bigr)
\Bigr)^2
\end{aligned}
\end{equation*}

Crucially, the focus of this study is to isolate and observe the impact of candidate document composition (sampling) strategies on distillation efficiency and generalization under identical objective functions. Implementation details are provided in Table~\ref{tab:setup}.

\subsection{Evaluation}
For in-domain evaluation, we select MS MARCO Dev and TREC Deep Learning (DL) Track 19~\cite{trec19}, as they share the same distribution as the training data. We follow the official evaluation protocol, reporting MRR@10 and Recall@1000 for MS MARCO Dev and nDCG@10 for TREC DL 19. For out-of-domain evaluation, we adopt the BEIR benchmark~\cite{thakur2021beir}. Following previous work~\cite{beir1,beir2,beir3}, we compute nDCG@10 across 13 datasets in BEIR, making the results directly comparable.

\section{EXPERIMENTAL RESULTS AND ANALYSIS}

Table~\ref{tab:main} quantitatively compares the impact of sampling strategies constructing the training data (retriever-top, low, mid, reranker-top, random, stratified) under three backbone models and two distillation objective functions (KLDiv, MarginMSE).

\subsection{Main Results}
The results in Table~\ref{tab:main} demonstrate that performance in distillation is governed 
% not by the average relevance of candidate documents itself, but 
by how wide a score range the candidate set within a query covers and how uniformly it is distributed. Specifically, random and stratified strategies consistently rank at the top across all backbones and objective functions, whereas strategies biased toward one end of the distribution, such as retriever-top, reranker-top, and low, cause performance degradation in many settings.
This trend is even more pronounced in out-of-domain evaluation. On BEIR-13, stratified achieves near-top performance across all three backbones in terms of both KL-divergence and MarginMSE (e.g., bert-base 0.314/0.318, co-condenser 0.365/0.376), and random also forms a strong baseline. Conversely, while reranker-top maintains moderate performance in some in-domain metrics, it exhibits instability in  
% settings involving domain shifts.
out-of-domain metrics.

From the perspective of objective functions, KL-divergence maintains relative rankings across strategies relatively stably, whereas MarginMSE leads to easy training collapse if sampling is inappropriate. For instance, a model trained with MarginMSE on reranker-top data using the co-condenser backbone virtually fails with MRR@10=.006, while stratified under the same conditions achieves MRR@10=.307. This suggests that regression-based objectives are more sensitive to distributional bias and noise in the negative set.

\begin{table}[h]
\centering
\small
\caption{Statistics of score-distribution diversity. We report the per-query mean of each metric.}
\label{tab:statistics}

\resizebox{\columnwidth}{!}{%
\begin{tabular}{l|ccc}
\toprule
\textbf{Sampling} & \textbf{Coverage} & \textbf{Entropy} & \textbf{Standard Deviation} \\
\midrule
stratified & \textbf{0.990} & \textbf{1.523} & \textbf{0.359} \\
random     & 0.759 & 1.270 & 0.295 \\
mid        & 0.323 & 0.902 & 0.139 \\
retriever-top   & 0.196 & 0.583 & 0.063 \\
reranker-top        & 0.106 & 0.202 & 0.035 \\
low        & 0.046 & 0.000 & 0.015 \\
\bottomrule
\end{tabular}%
}
\end{table}

% \begin{table}[h]
% \centering
% \small
% \caption{Statistics of score-distribution diversity. We report the per-query mean of each metric.}
% \label{tab:statistics}

% \resizebox{\columnwidth}{!}{%
% \begin{tabular}{l|cccccc}
% \toprule
% \textbf{Metric} & \textbf{stratified} & \textbf{random} & \textbf{mid} & \textbf{retriever-top} & \textbf{reranker-top} & \textbf{low} \\
% \midrule
% \textbf{Coverage} & \textbf{0.990} & 0.759 & 0.323 & 0.196 & 0.106 & 0.046 \\
% \textbf{Entropy} & \textbf{1.523} & 1.270 & 0.902 & 0.583 & 0.202 & 0.000 \\
% \textbf{Std} & \textbf{0.359} & 0.295 & 0.139 & 0.063 & 0.035 & 0.015 \\
% \bottomrule
% \end{tabular}%
% }
% \end{table}
\subsection{Diversity Statistics}

Table~\ref{tab:statistics} demonstrates how diversely the candidate document sets constructed by each sampling strategy cover the teacher's score distribution. We compute metrics for each query based on Min-max normalized scores. Specifically, Coverage is defined as the score range ($\max - \min$), Entropy as the Shannon entropy after dividing the score range into 8 equal-width bins, and Standard Deviation as the standard deviation of scores.

Experimental results show that Stratified Sampling records the highest values across all metrics ($Cov=0.990, Ent=1.523, Std=0.359$). This implies that the strategy does not bias toward specific score bands but evenly reflects the entire landscape of preferences assigned by the teacher in the training data. In contrast, widely used Top or Standard strategies show very low Entropy and Std, indicating that they only fragmentarily observe a tiny portion of the teacher's knowledge.

Importantly, the ranking of diversity metrics shown in Table~\ref{tab:statistics} largely aligns with the model performance ranking in Table~\ref{tab:main}. This strongly suggests that preserving the distribution defined by the teacher is far more critical for securing generalization performance than merely intensively training on hard negatives during distillation.

\begin{figure}[ht]
    \centering
    \includegraphics[width=\columnwidth]{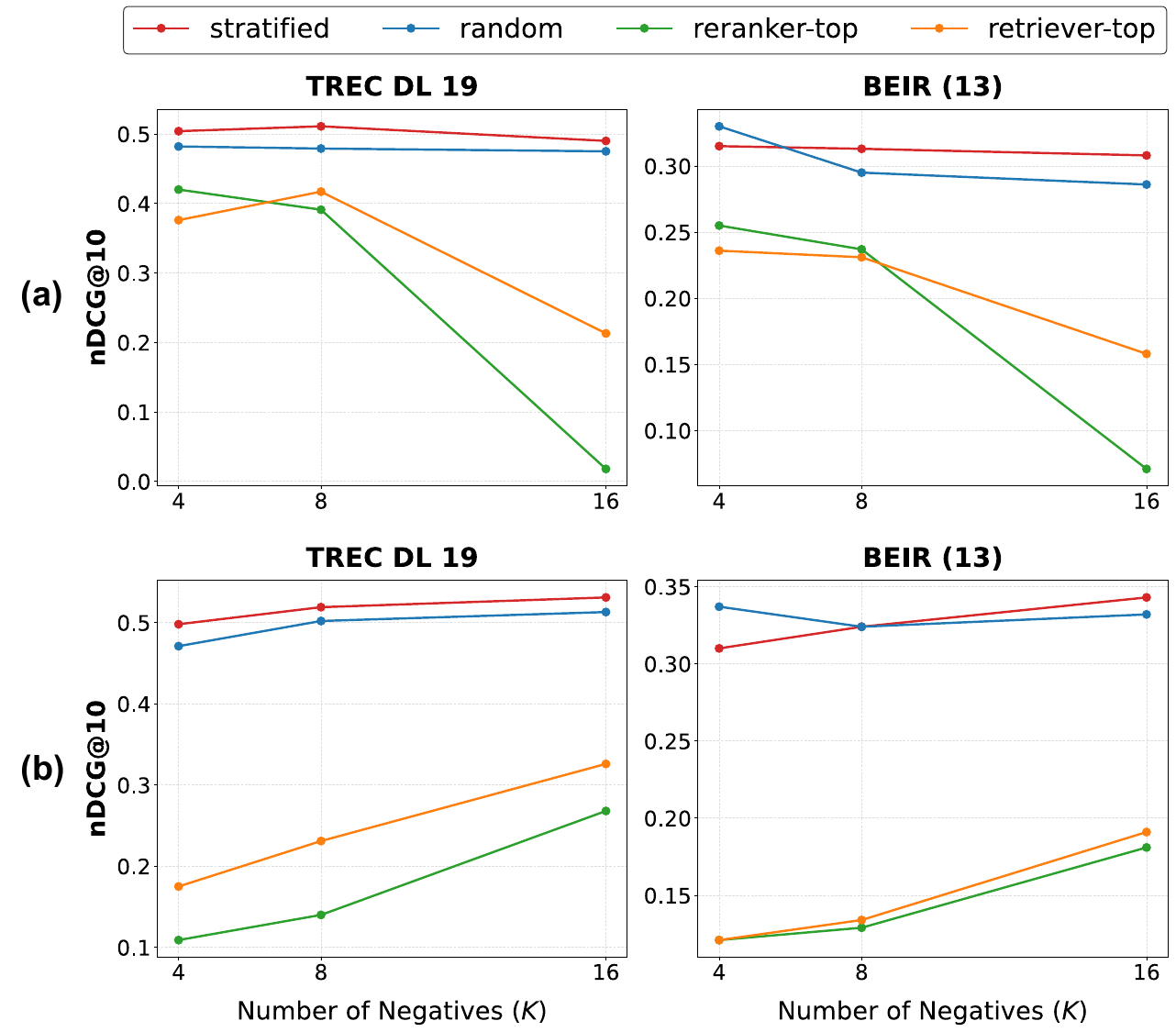}
    \caption{Retrieval performance (nDCG@10) on TREC DL 19 (in-domain) and BEIR (out-of-domain) as the number of sampled candidates $K$ varies ($K\in\{4,8,16\}$). Models are trained with KL-divergence in (a) and with MarginMSE in (b).}
    \label{fig:robustness}
\end{figure}

\subsection{Robustness of Stratified Sampling}
Figure~\ref{fig:robustness} presents the results comparing the impact of varying the number of sampled candidates $K$ on retrieval performance for four strategies (stratified, random, reranker-top, retriever-top). The experiment used the distilbert-base model as the backbone, applying (a) KL-Divergence and (b) MarginMSE as objective functions, respectively.

The most notable aspect of the results is the superior robustness of Stratified Sampling. The stratified strategy does not react sensitively to changes in $K$ values and consistently outperforms other strategies in almost all experimental settings. The only exception is when $K=4$ on the BEIR benchmark, where the Random strategy has a slight edge; this is interpreted as the Random method obtaining minimal diversity by chance covering the score distribution broadly when the initial $K$ is small.

However, as $K$ increases, the true value of stratified becomes apparent. As $K$ grows, the stratified method systematically fills gaps in the score range, providing mutually complementary learning signals rather than redundant difficulty levels. This leads to stable performance improvements unbiased toward specific difficulties, a trend observed in both In-domain (TREC DL 19) and Out-of-domain (BEIR). Notably, the highest performance was achieved with MarginMSE, $K=16$, and stratified settings (TREC DL 19: nDCG@10=0.531, BEIR: nDCG@10=0.343), suggesting that the representativeness of data distribution is key to generalization performance.

% \section{CONCLUSION}
% 본 연구는 Dense Retrieval 모델의 Knowledge Distillation(KD) 과정에서 간과되었던 학습 데이터의 Score Distribution이 모델의 성능과 일반화 능력에 미치는 영향을 심층적으로 분석힌다. 우리는 실험을 통해 기존의 관성적인 Hard Negative 중심 샘플링이 Teacher 모델의 풍부한 선호 정보를 온전히 전달하지 못함을 보였으며, 이를 해결하기 위해 점수 구간을 균등하게 커버하는 Stratified Sampling 전략을 제안한다.
% 벤치마크 실험 결과, Stratified Sampling은 in-domain(MS MARCO, TREC DL)뿐만 아니라 out-of-domain(BEIR) 환경에서도 기존의 Top-K 및 Random 방식 대비 일관되게 우수한 성능을 입증한다. 특히 샘플 수($K$)와 목적 함수(Loss Function)의 변화에도 robust한 성능을 유지하며, 데이터 효율적인 학습이 가능함을 확인한다.

% 결론적으로, 본 연구는 고품질의 retriever를 학습시키기 위해서는 단순히 어려운 오답을 구별하는 것을 넘어, Teacher가 정의하는 전역적인 랭킹 기하학(Global Ranking Geometry)을 보존하는 것이 필수적임을 시사한다. 우리가 제안한 데이터 구성 관점은 향후 효율적이고 일반화된 검색 모델 학습을 위한 새로운 표준이 될 것으로 기대한다.

\section{CONCLUSION}
This study provides an in-depth analysis of the impact of score distribution on the generalization capability of dense retrieval models within the Knowledge Distillation (KD) process. 
Through experiments designed to isolate the distributional effects from mining heuristics, we demonstrate that conventional sampling often fails to convey the rich preference information of the teacher model. 
To address this limitation, we propose Stratified Sampling, a deterministic strategy designed to uniformly cover the entire score spectrum.
Benchmark results confirm that Stratified Sampling consistently outperforms existing Top-K and Random methods in both in-domain and out-of-domain environments. 
Notably, it maintains robust performance across variations in sample size ($K$) and objective functions, establishing itself as a robust, parameter-free baseline for future distillation research.

% In conclusion, this study suggests that training high-quality retrievers requires more than simply distinguishing hard negatives; it is essential to preserve the Global Ranking Geometry defined by the teacher. We anticipate that our perspective on data composition will serve as a new standard for training efficient and generalized retrieval models in the future.

%%acknowledgements
% 중점, 스타트업, 생성형 AI (진실성)
\begin{acks}
This research was supported by Basic Science Research Program through the National Research Foundation of Korea(NRF) funded by the Ministry of Education(NRF-2021R1A6A1A03045425). This work was supported by the Commercialization Promotion Agency for R\&D Outcomes(COMPA) grant funded by the Korea government(Ministry of Science and ICT)(2710086166). This work was supported by Institute for Information \& communications Technology Promotion(IITP) grant funded by the Korea government(MSIT) 
(RS-2024-00398115, Research on the reliability and coherence of outcomes produced by Generative AI)
\end{acks}

\bibliographystyle{ACM-Reference-Format}
\bibliography{custom}

% \appendix

\end{document}